\documentclass[]{article}
\usepackage{graphicx}

\begin{document}

\title{Pseudoscalar Gluinonia to 750 GeV LHC Diphotons}
\author{C.T. Potter \\ Physics Department, University of Oregon}
\date{\today}

\maketitle

\abstract{Interpreting the diphoton mass excesses near 750~GeV reported by ATLAS and CMS as pseudoscalar gluinonia, bound states of two gluinos, we perform a scan in Next-to-Minimal Supersymmetry parameter space, fixing $m_{\tilde{g}}\approx 380$~GeV and identifying an experimentally viable point. We generate events with this point, perform fast simulation, and carry out an analysis modeled on the ATLAS search which reproduces features of the diphoton excess. This interpretation requires an enhancement of the signal strength above the nominal rate found in the literature and neglecting the gaugino mass unification constraint $M_{3}\approx 3M_{2}$.}

\section{Introduction}

The Standard Model (SM) of particle physics is both a success and a failure. A success because it is a strongly predictive model which no experimental measurement has falsified. A failure because it does not account for Dark Matter, the anomalous muon magnetic moment, the strong CP problem or the hierarchy problem. 

Supersymmetry (SUSY) can succeed where the SM fails and embed the SM as a low energy approximation, thus inheriting its successes \cite{Martin:1997ns}. The $\mu$-term problem of Minimal SUSY (MSSM) \cite{Martin:1997ns,Maniatis:2009re,Ellwanger:2009dp} motivates Next-to-Minimal SUSY (NMSSM), which includes a Higgs singlet in addition to the two Higgs doublets of the MSSM \cite{Maniatis:2009re,Ellwanger:2009dp}. In the study \cite{Potter:2015wsa} we define a natural  NMSSM benchmark $h_{60}$, characterized by an effective MSSM and a slightly broken PQ symmetry. It features a light pseudoscaler Higgs with $m_{a_1}\approx10$~GeV and a light scalar with $m_{h_1} \approx 60$~GeV, which can be produced from cascade decay of a light stop with $m_{\tilde{t}_{1}} \approx 340$~GeV to electroweakinos $\tilde{t}_{1} \rightarrow \chi^{+}b \rightarrow \chi_{3}Wb$, and $\chi_{3} \rightarrow h_1 \chi_1 \rightarrow 2a_1 \chi_1$.

ATLAS and CMS have reported possible excesses in the diphoton mass spectrum \cite{CMS-PAS-EXO-15-004,ATLAS-CONF-2015-081,ATLAS-CONF-2016-018} at the the Large Hadron Collider. 
ATLAS reports a local significance of 3.6$\sigma$ near 750 GeV, while CMS reports a local significance of 2.6$\sigma$ near 760 GeV.   A diphoton decay strongly indicates decay from either a scalar or pseudoscalar. ATLAS reports a fitted width of approximately  40~GeV for the 750 GeV feature, much larger than the diphoton energy resolution. Assuming a resonance decay, the width is unusually large. Alternatively, there may be two or more resonances in close proximity. 

See \cite{Strumia:2016wys} for a recent bibliography of several hundred interpretations of the diphoton excess. In this paper we interpret the diphoton excess as gluinonia, bound states of gluinos which have been dubbed the hydrogen atom of SUSY \cite{GOLDMAN1985181}.

\section{Gluinonia}

The benchmark $h_{60}$ in \cite{Potter:2015wsa} features a relatively light gluino with $m_{\tilde{g}} \approx 610$~GeV which has been excluded by ATLAS and CMS in the $\tilde{g} \rightarrow t \tilde{t}_{1}$ channel. Since the lower energy phenomenology can be decoupled by relaxing the gaugino mass unification constraint $M_{3} \approx 3M_{2}$ imposed during the NMSSMTools scan, that study simply assumes the gluino mass is higher than the constraint imposes and thus avoids exclusion due to a lower gluino pair production cross section.

Alternatively, the gluino may have a small enough width that it can form bound states with shorter lifetimes than the gluino itself.  We consider the possibility that the diphoton excess reported by ATLAS and CMS is a bound state of two gluinos, in which case the gluino mass is $m_{\tilde{g}} \approx 380$~GeV. This hypothesis is amenable to forming bound states since this gluino is below threshold for $\tilde{g} \rightarrow t \tilde{t}_1$ if $m_{\tilde{t}}> 207$~GeV, forcing decays through virtual quarks to electroweakinos thereby increasing the gluino lifetime. In this case, pair produced gluinos may form states bound by the strong interaction. Such gluinonia states, the so-called hydrogen atom of SUSY, have been studied in the literature as early as the 1980s \cite{Keung:1983wz,GOLDMAN1985181,Cheung:2004ad,Kauth:2009ud,Hagiwara:2009hq,Kats:2009bv,Kats:2016kuz}.

In this study we consider a natural NMSSM benchmark $\tilde{g}_{380}$ with the low energy phenomenology of $h_{60}$ but with a different spectrum for the gluino, stop and sbottom. We take the 750~GeV LHC diphoton excess to be two or more pseudoscalar gluinonia states, and the gluino mass to be $m_{\tilde{g}} \approx 380$~GeV.

We consider the the ground state color singlet $1^{1}S_{0}(1)$ (hereafter $G_1$) and its $n$th radial excitations $n^{1}S_{0}(1)$ (hereafter $G_n$). This state has been studied in the literature \cite{Cheung:2004ad,Kauth:2009ud,Hagiwara:2009hq,Kats:2009bv} with consistent results, though there is considerable uncertainty in the evaluation of the ground state wavefunction at the origin $\vert \psi_{0}(0))\vert^2$ which induces uncertainty in the cross section and width. For $m_{\tilde{g}} \approx 380$~GeV, the width is $\Gamma_1 \approx 320$~MeV. The binding energy of the ground state is $E_{b}^{1} \approx 20$~GeV, while for the $n$th radial excitations the energy is $E_{b}^{n} =  E_{b}^{1}/n^2$. 

The difference in binding energies, together with detector resolution, can account for the large fitted width of approximately 40~GeV reported by ATLAS since the ground state and excitations form a spectrum with  masses $2M_{\tilde{g}}-E_{b}^{n}$. The LHC production cross section at $\sqrt{s}=14$~TeV is $\sigma_1  \approx 4$~pb using the narrow width approximation. For the $n$th radial excitation, the Coulomb radial wavefunction scales like $n^{-3/2}$, and

\begin{eqnarray}
\frac{\sigma_n}{\sigma_1} & \approx & \frac{\vert \psi_{n}(0) \vert^{2}}{\vert \psi_{1}(0) \vert^{2}} \\
& \approx & n^{-3} 
\end{eqnarray}

\noindent where the small mass differences have been negelected. The total cross section for the ground state and all radial excitations is $\sigma = \sum_{n=1}^{\infty} \sigma_1/n^3 =\zeta(3) \sigma_1 \approx 1.2\sigma_1$.

The digluon width $\Gamma_{gg}$ dominates the total width $\Gamma_1$. Diphoton decays of these states are suppressed relative to digluon decays by $\alpha_{em}^2$, with $R_{\gamma \gamma}=\Gamma_{\gamma \gamma}/\Gamma_{gg} \approx 5 \times  10^{-5}$ \cite{Kauth:2009ud}, but we consider that some mechanism has enhanced production of the diphoton final state either through additional gluinonia states, enhanced cross section, enhanced branching ratios, or some combination of these possibilities. The nominal signal strength for $pp \rightarrow G_1 \rightarrow \gamma \gamma$ is $\sigma_{1} \times R_{\gamma \gamma} \approx0.2$~fb, requiring an enhancement of order $\times 40$ to produce the LHC 750 GeV diphoton excess if a signal efficiency of 40\% is assumed. Throughout this paper we assume an enhanced cross section $4 \sigma_1$ for $pp \rightarrow G_1$ and assume the residual enhancement $\times 10$ arises from some other effect.

\section{NMSSM Benchmark $\tilde{g}_{380}$}

\begin{table}[t!]
\begin{center}
\begin{tabular}{|c|c|c|c|} \hline
Parameter & Scan Range & $\tilde{g}_{380}$  & $h_{60}$\\ \hline \hline
 $\lambda$ & Fixed & 0.03505 & 0.03505 \\
 $\kappa$ & Fixed & 0.006088 &0.006088 \\
 $m_{A}$ & Fixed &  1068.~GeV & 1068.~GeV\\
  $m_{P}$ & Fixed & 10.25~GeV  & 10.25~GeV\\
  $\mu_{eff}$ & [150,180]~GeV &   173.5~GeV & 166.7~GeV \\ 
  $\tan \beta$ & [1,25] & 6.01 &   15.49  \\ \hline
 $M_{1}$ & Fixed & 80.73~GeV &   80.73~GeV \\
 $M_{2}$ & Fixed & 161.5~GeV  & 161.5~GeV\\
 $M_{3}$ & Fixed & 280.0~GeV & 484.4~GeV \\ \hline
 $X_{t}$ & [$0,2X_{t}^{max}$] & 1282.~GeV &  1378.~GeV\\
 $m_{\tilde{Q3}_{L}}$ & [500,600]~GeV & 546.3~GeV  &  546.9~GeV\\
 $m_{\tilde{U3}_{R}}$ & $m_{\tilde{Q3}_{L}}$ & 546.3~GeV  &  546.9~GeV\\ \hline 
\end{tabular}
\caption{NMSSMTools parameter ranges with their values for benchmarks $\tilde{g}_{380}$ and $h_{60}$.} 
\label{tab:scan}
\end{center}
\end{table}

We perform a NMSSM scan similar to the scan which produced $h_{60}$. We use NMSSMTools4.8.2 \cite{Ellwanger:2004xm,Ellwanger:2005dv,Belanger:2005kh,Ellwanger:2006rn,Das:2011dg,Muhlleitner:2003vg} and impose the full set of experimental constraints. Unlike the $h_{60}$ scan, however, we use Higgs mass precision 1 rather than 2 in order to expedite the scan.

We fix the doublet-singlet coupling $\lambda$, the singlet self interaction coupling $\kappa$, the doublet scalar mass $m_{A}$, and the doublet pseudoscalar mass $m_{P}$ to their $h_{60}$ values.  For $\tilde{g}_{380}$, we target a $380$~GeV gluino by relaxing the unification constraint for the gaugino masses $M_3=3M_2$ and directly fix the mass $M_{3}=280$~GeV. The other gaugino masses are fixed to the $h_{60}$ values $M_{2}=161.5$~GeV and $M_{1}=\frac{1}{2}M_{2}$.

We scan four parameters close to their $h_{60}$ values with $10^{8}$ random points:  third generation squark mass $m_{Q_3}$, stop mixing $X_{t}$, the  effective $\mu$-term $\mu_{eff}$, and the ratio of doublet VEVs $\tan \beta$. All other squark and soft trilinear parameters are fixed to 1500~GeV. The slepton mass parameters are fixed to 300~GeV.  See Table \ref{tab:scan} for the scan parameter values and ranges. 

Since, as noted in \cite{Potter:2015wsa}, a stop lighter  than the one in $h_{60}$ can explain the CMS dilepton excess \cite{Khachatryan:2015lwa}, we seek the benchmark $\tilde{g}_{380}$ among the points surviving the scan constraints with the lowest stop mass which is still consistent with the lower energy phenomenology of $h_{60}$. These criteria yield the point $\tilde{g}_{380}$ with $m_{\tilde{g}}=381.7$~GeV, $m_{\tilde{t}_1}=324.0$~GeV and $m_{\tilde{b}_1}=527.8$~GeV. The $\tilde{g}_{380}$ gluino width is $\Gamma_{\tilde{g}} \approx 2$~MeV, which,  comparing with the nominal gluinonum width $\Gamma_{1} \approx 320$~MeV, assures that the gluino lifetime is sufficiently long to allow gluinonium production and decay. In the assumed case here, with enhanced production cross section $4 \sigma_1$, the width is even larger, $\Gamma_{G_1} \approx 1.3$~GeV. See Table \ref{tab:scan} for the parameter values in $\tilde{g}_{380}$ and, for comparison, $h_{60}$.

For verification that pair production in $\tilde{g}_{380}$ of neutralinos, charginos, stops, and gluinonium decaying to to digluons is allowed given LHC8 constraints, we use CheckMATE \cite{Drees:2013wra}. See Table \ref{tab:checkmate} for the maximum exclusion $r_{max}$ from all validated ATLAS and CMS analyses. A process is excluded if $r_{max}>1$. The table assumes a digluon signal strenth $\sigma_{pp \rightarrow G_1} \times BR(G_1 \rightarrow gg)=4\sigma_{1} \approx 16$~pb. 

For the $pp \rightarrow \tilde{g} \tilde{g}$ processes with bare gluinos decaying before forming bound states, the cross section is varied from the nominal Pythia8 cross section for $m_{\tilde{g}}=380$~GeV of $\sigma_{pp \rightarrow \tilde{g} \tilde{g}} \approx 16$~pb  $\times 1, \times \frac{1}{16}, \times \frac{1}{32}$ in the CheckMATE test. The results indicate that $\sigma_{pp \rightarrow \tilde{g} \tilde{g}} \approx 0.5$~pb without bound state formation is allowed but $\sigma_{pp \rightarrow \tilde{g} \tilde{g}} \approx 1.0$~pb is excluded.

\section{LHC Diphoton Signature of $\tilde{g}_{380}$}

In the previous section, we used Pythia8.205 \cite{Sjostrand:2007gs,Sjostrand:2006za} to simulate  gluon fusion production and digluon decay $gg \rightarrow G_1 \rightarrow gg$ for the CheckMATE test. In this section, we use the same generator for gluon fusion production and diphoton decay $gg \rightarrow G_1 \rightarrow \gamma \gamma$. For the diphoton background simulation, we use MG5\_aMC@NLO \cite{Alwall:2014hca} to simulate $qq \rightarrow \gamma \gamma$, $qg \rightarrow q \gamma \gamma$ and the box process $gg \rightarrow \gamma \gamma$.

The center of mass energy is set to $\sqrt{s}=13$~TeV with $pp$ beams. Gluon fusion production of the MSSM pseudoscalar Higgs $A$ is employed to mimic pseudoscalar gluinonia production. The decay $A \rightarrow \gamma \gamma$ is required, and two masses are specified: $m_{A}=740$~GeV to mimic the $G_1 \rightarrow \gamma \gamma$ decay and $m_{A}=760$~GeV to mimic the diphoton decay of the remaining radial excitations $G_n \rightarrow \gamma \gamma$. The width is set to be negligible compared to the detector resolution.

Fast detector simulation is carried out with Delphe3.2.0 \cite{deFavereau:2013fsa} using the Delphes3 ATLAS card with pileup suitable for $\sqrt{s}=13$~TeV. The ATLAS diphoton search selection is reproduced as far as this is possible with fast simulation.  Photons are required to satisfy $\vert \eta_{\gamma} \vert < 2.37$, excluding the region $1.37 < \vert \eta_{\gamma} \vert < 1.52$. Photon isolation requires $E_{cal}^{0.4}/E_{\gamma}<0.022$ where $E_{cal}^{0.4}$ is the calorimeter energy in a cone of radius $\Delta R=0.4$  around (but excluding) the photon. The analysis requirements are these:

\begin{itemize}

\item at least two isolated photons with $E_{T}^{\gamma}>30$~GeV 

\item at least one isolated photon with $E_{T}^{\gamma}>40$~GeV

\item leading photon satisfies $E_{T}^{\gamma}/m_{\gamma \gamma}>0.4$

\item subleading photon satisfies $E_{T}^{\gamma}/m_{\gamma \gamma}>0.3$

\end{itemize}

\noindent where $m_{\gamma \gamma}$ is the diphoton mass. After full signal selection, the signal efficiency is approximately 40\% for both signal samples.

\begin{figure}[t]
\begin{center}
\includegraphics[width=2.8in]{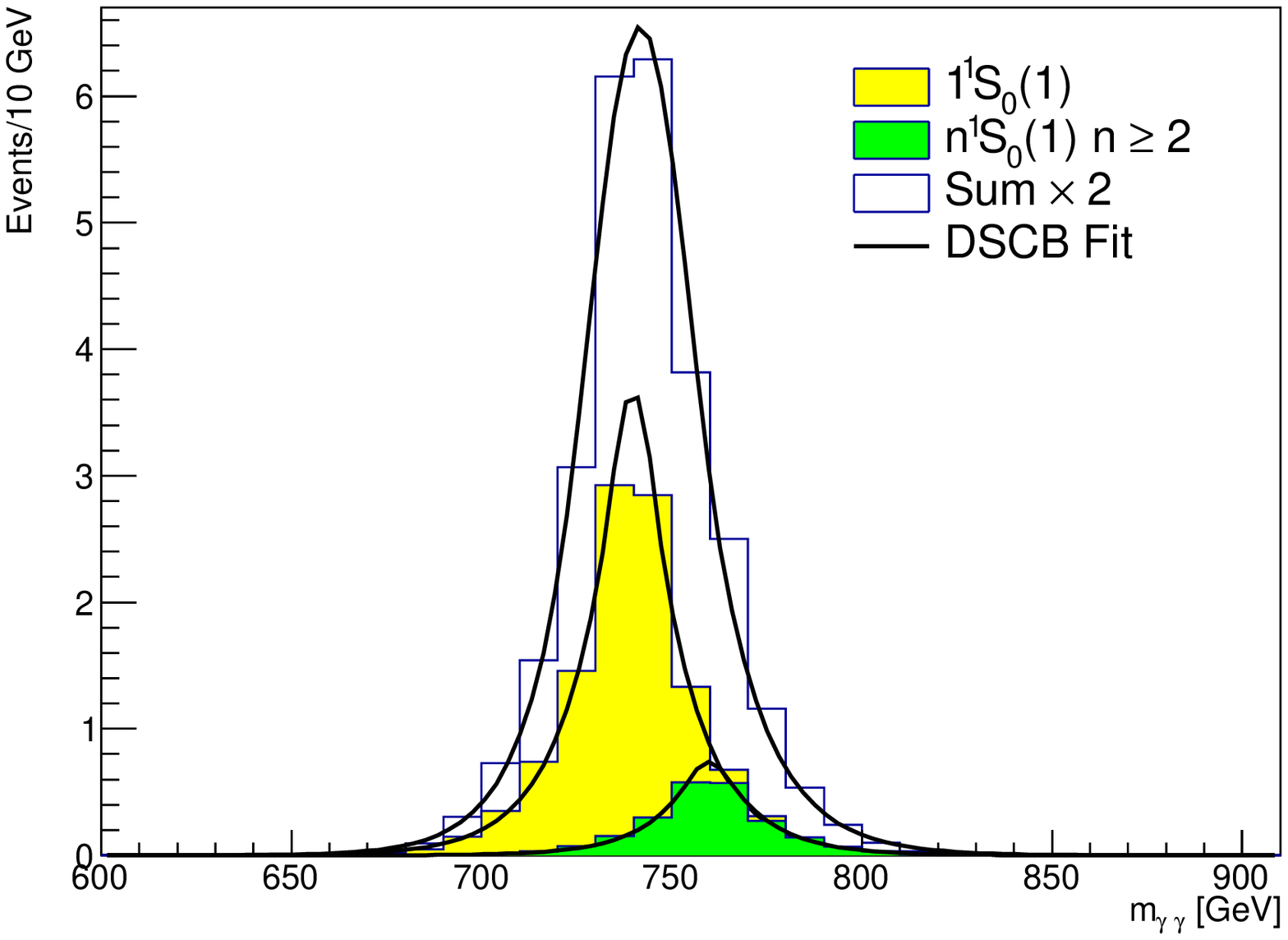}
\includegraphics[width=2.8in]{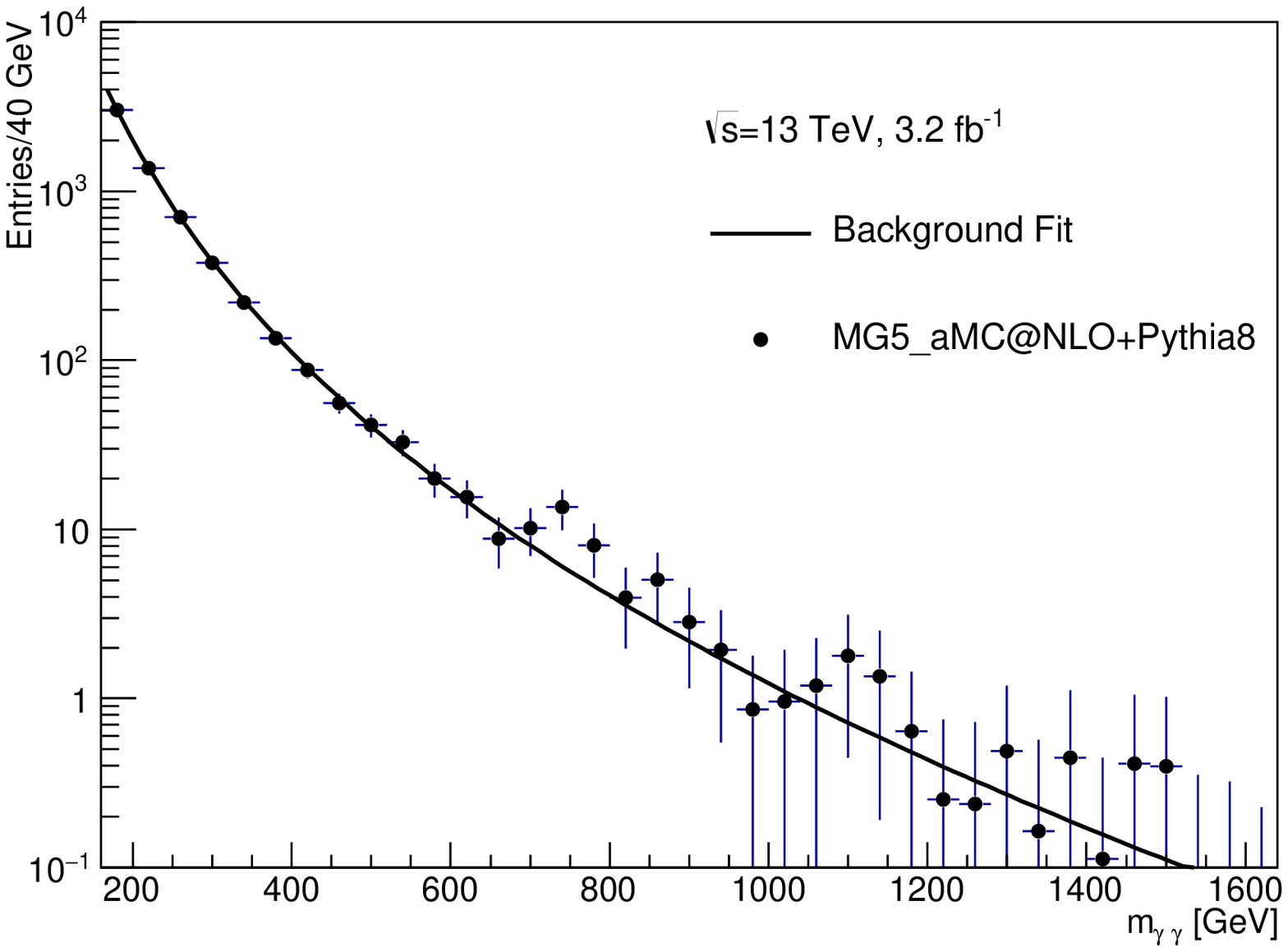}
\caption{Signal (left) and signal plus fitted background (right) after full signal selection. }
\label{fig:diphoton}
\end{center}
\end{figure}

See Figure~\ref{fig:diphoton} for the signal diphoton spectrum after full signal selection, assuming $\sigma_{pp \rightarrow G_1} \times BR(G_1 \rightarrow \gamma \gamma)=8$~fb, $\sqrt{s}=13$~TeV and $\int dt \mathcal{L}=3.2$~fb$^{-1}$. The $m_{\gamma \gamma}$ distributions for the $G_1$, the radial excitations $G_n$, and their sum are fit with a double sided Crystal Ball (DSCB) as defined in the ATLAS search \cite{ATLAS-CONF-2015-081}. The fitted full width at half maximum is approximately 40~GeV. 

Also shown in Figure~\ref{fig:diphoton} is the diphoton spectrum of the Pythia8 signal added to the MG5\_aMC@NLO background  processes $q q \rightarrow \gamma \gamma$, $qg \rightarrow q \gamma \gamma $ and $gg \rightarrow \gamma \gamma$ together with a background fit. The fit employs the probability density function used by ATLAS, namely $f(x;b,a_{0})=(1-x^{1/3})^{b} x^{a_0}$ where $x=m_{\gamma \gamma}/\sqrt{s}$ and $a_0,b$ are free parameters \cite{ATLAS-CONF-2015-081}.
.
\begin{table}[t]
\begin{center}
\begin{tabular}{|l|c|c|c|c|} \hline
Process & $r_{max}^{ATLAS}$ & Analysis & $r_{max}^{CMS}$& Analysis \\ \hline
$pp \rightarrow \chi \chi$ & 0.37 & atlas\_conf\_2013\_035 & 0.08 & cms\_1303\_2985 \\
$pp \rightarrow \tilde{t}_{1} \tilde{t_{1}^{\star}}$ & 0.15 & atlas\_conf\_2013\_047 & 0.44 & cms\_1502\_06031\\
$pp \rightarrow G_1 \rightarrow gg$ & 0.65 & atlas\_1308\_1841 &  0.00 & cms\_1303\_2985 \\ \hline
$pp \rightarrow \tilde{g} \tilde{g} \times 1$ & 3.33 & atlas\_conf\_2013\_089 & 2.89 & cms\_1502\_06031\\ 
$pp \rightarrow \tilde{g} \tilde{g} \times \frac{1}{16}$ & 1.31 & atlas\_1308\_1841 & 0.77 & cms\_1502\_06031\\ 
$pp \rightarrow \tilde{g} \tilde{g} \times \frac{1}{32}$ &0.62 &  atlas\_1308\_1841 & 0.38 & cms\_1502\_06031 \\ \hline
\end{tabular}
\caption{Exclusion $r_{max}$ by ATLAS and CMS analyses, obtained with CheckMATE.}
\label{tab:checkmate}
\end{center}
\end{table}

\section{Conclusion}

We have identified pseudoscalar gluinonia, bound states of gluinos with $m_{\tilde{g}} \approx 380$~GeV, as an explanation for the diphoton excesses reported by ATLAS and CMS. A light gluino below the threshold for decay to stop or sbottom has a small width because it must decay via virtual squarks, allowing two gluinos to form bound states with widths much larger than the gluino width. 

A scan is performed in NMSSM parameter space with NMSSMTools4 to identify a benchmark point $\tilde{g}_{380}$ consistent with the benchmark $h_{60}$ \cite{Potter:2015wsa} featuring  $m_{\tilde{g}} \approx 380$~GeV  which survives experimental constraints. In order to identify a viable point with a gluino mass small enough to lie below the threshold for two-body decay, the gaugino mass unification constraint $M_3 \approx 3M_2$ is ignored. We verify that $\tilde{g}_{380}$ survives LHC8 search constraints with CheckMATE.

An analysis based on the ATLAS diphoton search is carried out with events generated by Pythia8 with Delphes3 detector simulation which reproduces the important features of the diphoton excess in the $\sqrt{s}=13$~TeV data. We find that to reproduce the 750 GeV diphoton excess requires an enhancement of the nominal gluinonium to diphoton signal strength of $\times 40$, and assume that some of this enhancement is due to the underestimate of the ground state wavefunction at the origin $\vert \psi(0)\vert^2$ in the literature. An explanation for the remaining enhancement is undetermined.

\begin{center}\textbf{Acknowledgements}\end{center}
The author thanks Yevgeny Kats for pointing out an elementary problem with the first draft of this paper and the Alder Institute for High Energy Physics for financial support. 

\bibliography{paper}

\begin{thebibliography}{10}

\bibitem{Martin:1997ns}
Stephen~P. Martin.
\newblock {A Supersymmetry primer}.
\newblock 1997, hep-ph/9709356.

\bibitem{Maniatis:2009re}
M.~Maniatis.
\newblock {The Next-to-Minimal Supersymmetric extension of the Standard Model
  reviewed}.
\newblock {\em Int.J.Mod.Phys.}, A25:3505--3602, 2010, 0906.0777.

\bibitem{Ellwanger:2009dp}
Ulrich Ellwanger, Cyril Hugonie, and Ana~M. Teixeira.
\newblock {The Next-to-Minimal Supersymmetric Standard Model}.
\newblock {\em Phys.Rept.}, 496:1--77, 2010, 0910.1785.

\bibitem{Potter:2015wsa}
C.~T. Potter.
\newblock {Natural NMSSM with a Light Singlet Higgs and Singlino LSP}.
\newblock {\em Eur. Phys. J.}, C76(1):44, 2016, 1505.05554.

\bibitem{CMS-PAS-EXO-15-004}
{Search for new physics in high mass diphoton events in proton-proton
  collisions at $\sqrt{s} = 13$ TeV}.
\newblock Technical Report CMS-PAS-EXO-15-004, CERN, Geneva, 2015.

\bibitem{ATLAS-CONF-2015-081}
{Search for resonances decaying to photon pairs in 3.2 fb$^{-1}$ of $pp$
  collisions at $\sqrt{s}$ = 13 TeV with the ATLAS detector}.
\newblock Technical Report ATLAS-CONF-2015-081, CERN, Geneva, Dec 2015.

\bibitem{ATLAS-CONF-2016-018}
{Search for resonances in diphoton events with the ATLAS detector at $\sqrt{s}$
  = 13 TeV}.
\newblock Technical Report ATLAS-CONF-2016-018, CERN, Geneva, Mar 2016.

\bibitem{Strumia:2016wys}
Alessandro Strumia.
\newblock {Interpreting the 750 GeV digamma excess: a review}.
\newblock 2016, 1605.09401.

\bibitem{GOLDMAN1985181}
T.~Goldman and H.E. Haber.
\newblock Gluinonium: The hydrogen atom of supersymmetry.
\newblock {\em Physica D: Nonlinear Phenomena}, 15(1):181 -- 196, 1985.

\bibitem{Keung:1983wz}
Wai-Yee Keung and Avinash Khare.
\newblock {GLUINOBALLS}.
\newblock {\em Phys. Rev.}, D29:2657, 1984.

\bibitem{Cheung:2004ad}
Kingman Cheung and Wai-Yee Keung.
\newblock {Split supersymmetry, stable gluino, and gluinonium}.
\newblock {\em Phys. Rev.}, D71:015015, 2005, hep-ph/0408335.

\bibitem{Kauth:2009ud}
Matthias~R. Kauth, Johann~H. Kuhn, Peter Marquard, and Matthias Steinhauser.
\newblock {Gluinonia: Energy Levels, Production and Decay}.
\newblock {\em Nucl. Phys.}, B831:285--305, 2010, 0910.2612.

\bibitem{Hagiwara:2009hq}
Kaoru Hagiwara and Hiroshi Yokoya.
\newblock {Bound-state effects on gluino-pair production at hadron colliders}.
\newblock {\em JHEP}, 10:049, 2009, 0909.3204.

\bibitem{Kats:2009bv}
Yevgeny Kats and Matthew~D. Schwartz.
\newblock {Annihilation decays of bound states at the LHC}.
\newblock {\em JHEP}, 04:016, 2010, 0912.0526.

\bibitem{Kats:2016kuz}
Yevgeny Kats and Matthew~J. Strassler.
\newblock {Resonances from QCD bound states and the 750 GeV diphoton excess}.
\newblock {\em JHEP}, 05:092, 2016, 1602.08819.

\bibitem{Ellwanger:2004xm}
Ulrich Ellwanger, John~F. Gunion, and Cyril Hugonie.
\newblock {NMHDECAY: A Fortran code for the Higgs masses, couplings and decay
  widths in the NMSSM}.
\newblock {\em JHEP}, 0502:066, 2005, hep-ph/0406215.

\bibitem{Ellwanger:2005dv}
Ulrich Ellwanger and Cyril Hugonie.
\newblock {NMHDECAY 2.0: An Updated program for sparticle masses, Higgs masses,
  couplings and decay widths in the NMSSM}.
\newblock {\em Comput.Phys.Commun.}, 175:290--303, 2006, hep-ph/0508022.

\bibitem{Belanger:2005kh}
G.~Belanger, F.~Boudjema, C.~Hugonie, A.~Pukhov, and A.~Semenov.
\newblock {Relic density of dark matter in the NMSSM}.
\newblock {\em JCAP}, 0509:001, 2005, hep-ph/0505142.

\bibitem{Ellwanger:2006rn}
Ulrich Ellwanger and Cyril Hugonie.
\newblock {NMSPEC: A Fortran code for the sparticle and Higgs masses in the
  NMSSM with GUT scale boundary conditions}.
\newblock {\em Comput.Phys.Commun.}, 177:399--407, 2007, hep-ph/0612134.

\bibitem{Das:2011dg}
Debottam Das, Ulrich Ellwanger, and Ana~M. Teixeira.
\newblock {NMSDECAY: A Fortran Code for Supersymmetric Particle Decays in the
  Next-to-Minimal Supersymmetric Standard Model}.
\newblock {\em Comput.Phys.Commun.}, 183:774--779, 2012, 1106.5633.

\bibitem{Muhlleitner:2003vg}
M.~Muhlleitner, A.~Djouadi, and Y.~Mambrini.
\newblock {SDECAY: A Fortran code for the decays of the supersymmetric
  particles in the MSSM}.
\newblock {\em Comput.Phys.Commun.}, 168:46--70, 2005, hep-ph/0311167.

\bibitem{Khachatryan:2015lwa}
Vardan Khachatryan et~al.
\newblock {Search for Physics Beyond the Standard Model in Events with Two
  Leptons, Jets, and Missing Transverse Momentum in pp Collisions at sqrt(s) =
  8 TeV}.
\newblock {\em JHEP}, 04:124, 2015, 1502.06031.

\bibitem{Drees:2013wra}
Manuel Drees, Herbi Dreiner, Daniel Schmeier, Jamie Tattersall, and Jong~Soo
  Kim.
\newblock {CheckMATE: Confronting your Favourite New Physics Model with LHC
  Data}.
\newblock {\em Comput. Phys. Commun.}, 187:227--265, 2014, 1312.2591.

\bibitem{Sjostrand:2007gs}
Torbjorn Sjostrand, Stephen Mrenna, and Peter~Z. Skands.
\newblock {A Brief Introduction to PYTHIA 8.1}.
\newblock {\em Comput.Phys.Commun.}, 178:852--867, 2008, 0710.3820.

\bibitem{Sjostrand:2006za}
Torbjorn Sjostrand, Stephen Mrenna, and Peter~Z. Skands.
\newblock {PYTHIA 6.4 Physics and Manual}.
\newblock {\em JHEP}, 0605:026, 2006, hep-ph/0603175.

\bibitem{Alwall:2014hca}
J.~Alwall, R.~Frederix, S.~Frixione, V.~Hirschi, F.~Maltoni, O.~Mattelaer,
  H.~S. Shao, T.~Stelzer, P.~Torrielli, and M.~Zaro.
\newblock {The automated computation of tree-level and next-to-leading order
  differential cross sections, and their matching to parton shower
  simulations}.
\newblock {\em JHEP}, 07:079, 2014, 1405.0301.

\bibitem{deFavereau:2013fsa}
J.~de~Favereau, C.~Delaere, P.~Demin, A.~Giammanco, V.~Lemaître, A.~Mertens,
  and M.~Selvaggi.
\newblock {DELPHES 3, A modular framework for fast simulation of a generic
  collider experiment}.
\newblock {\em JHEP}, 02:057, 2014, 1307.6346.

\end{thebibliography}

\end{document}